\documentclass{aa}
\usepackage{graphics}

\begin{document}

   \thesaurus{01     
              (02.01.4;  
               02.18.5;  
               08.05.1;  
               08.09.2 HD44179;  
               08.09.2 RR Tel;  
               09.12.1)} 
   \title{H-atom laser without inversion (LWI) in space:  a possible explanation for
the intense, narrow-band, H($\alpha $) emission frequently observed in
reddened early-type stars}


   \author{P. P. Sorokin \and J. H. Glownia}

   \institute{IBM Research Division, P. O. Box 218, Yorktown Heights, NY 10598, USA (sorokin@us.ibm.com, glownia@us.ibm.com)} 
   \date{Received 6 June 2001; accepted}

    \titlerunning{H-atom laser without inversion (LWI) in space}
    \authorrunning{P. P. Sorokin \& J. H. Glownia}

   \maketitle

   \begin{abstract}

A model is suggested to explain the frequently observed presence of intense,
narrow-band, H($\alpha $) emission lines in the optical spectra of reddened,
early-type stars (e.g. HD 44179, IRAS 18179-1346, IRAS 20298+4011). It is
proposed that hydrogen atoms surrounding compact H II regions enveloping
such stars become coherently phased via a nonlinear photonic mechanism that
leads to `electromagnetically induced transparency (EIT)'. EIT is a powerful
technique that can be used to make a material system transparent to resonant
laser radiation, while still allowing large nonlinear resonant processes to
occur with high probability. In EIT terminology, a `$\Lambda $'
configuration, involving H-atom levels (1s, 3p, and 2s), is here assumed to
be operative. The EIT `coupling beam' is the narrow-band H($\alpha $)
radiation predicted to be coherently generated via a standard `laser without
inversion (LWI)' scenario when coherently phased atoms are excited to the 3p
level by means of a separate nonlinear excitation process known as resonant
hyper-Raman scattering (HRS). In the unit HRS pumping process, a pair of
far-ultraviolet (FUV) photons, with frequencies lying very close to Ly-$%
\beta $ but offset from it by equal amounts to high and low energies, are
absorbed from the star's blackbody continuum, a photon at Ly-$\beta $ line
center is emitted, and an atom is excited to the 3p level - with all events
in this energy conserving process occurring simultaneously. The EIT `probe
beam' is the light predicted to be coherently generated at Ly-$\beta $ line
center, which - as a result of the complete linear transparency afforded at
this frequency by the coherently phased H atoms - can propagate completely
unattenuated through the optically thick H-atom cloud surrounding the star.

   \end{abstract}

\section{Introduction}
The introduction of an entirely new concept to explain a class of observed
phenomena in any field of science would seem to be a redundant activity, if
the phenomena can reasonably be accounted for with use of existing theories
and models. However, in the cases of certain astronomical observations of H($%
\alpha $) emission, standard explanations appear to be inadequate. This
letter focusses specifically on recorded observations of intense H($\alpha $%
) emission in reddened early-type stars. It will be shown below that enough
spectral information exists for such stars to suggest strongly that a new
paradigm is here needed. One that appears to be quite suitable for
explaining the H($\alpha $) emission data was recently developed in the
field of quantum electronics and is termed `laser without inversion (LWI)'.
The LWI concept is itself inseparably tied to another recently developed
quantum electronics concept, that of  `electromagnetically induced
transparency (EIT)'. In the present letter a novel astrophysical model based
upon LWI/EIT is introduced that appears to be fully capable of explaining
the most intense and spectrally narrowest H($\alpha $) emissions seen in
reddened early-type stars.

\section{H($\alpha $) emissions in reddened early-type stars}

\begin{figure}
\resizebox{\hsize}{!}{\includegraphics{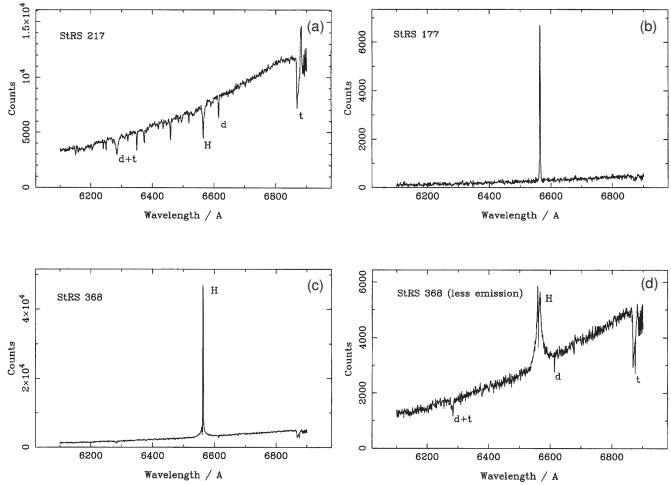}}
\caption{a-d. Spectra of three reddened early-type stars selected from Fig.~\ref{fig4}
of Rawlings et al. (\cite{rawlings}). Features labeled `d' and `t' are diffuse
interstellar and telluric features, respectively. Stars labeled by running
numbers used in Table 1 of Stephenson (\cite{stephenson}). (StRS $\equiv $ `Stephenson
Reddened Star'.) Standard identifications for StRS 177 and StRS 368 are IRAS
18179-1346 and IRAS 20298+4011, respectively.}
\label{fig1}
\end{figure}

In Rawlings et al. (\cite{rawlings}) new optical spectroscopy for 45\% of an older
catalog (Stephenson \cite{stephenson}) of $\sim $440 reddened stars was performed in an
attempt to isolate reddened stars which are also early-type. The spectra
between 6100 \AA\ and 6900 \AA\ of all the stars studied are shown in
Rawlings et al. (\cite{rawlings}). Only fifteen of the stars surveyed could be
classified as early-type stars. With regard to H($\alpha $), the spectra of
these early-type stars show tremendous variations. Some (Fig.~\ref{fig1}a) show H($%
\alpha $) only in absorption. Others (Figs.~\ref{fig1}b, ~\ref{fig1}c, ~\ref{fig1}d) show H($\alpha $)
strongly in emission, with an intensity completely disproportionate to all
other absorption features present. Yet the spectral types of these stars are
all apparently roughly comparable. In the spectra of all the cooler stars
surveyed by Rawlings et al. (\cite{rawlings}), H($\alpha $), if present at all, appears
only in absorption.

\begin{figure}
\resizebox{\hsize}{!}{\includegraphics{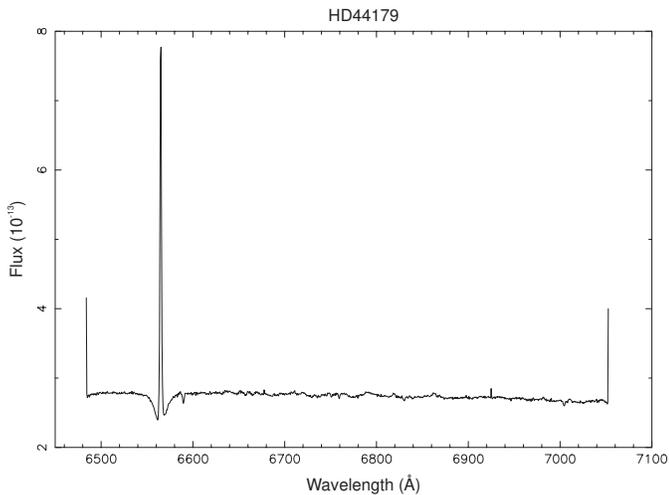}}
\caption{STIS scan of HD 44179 taken on Mar 26, 1998 (T. Gull, principal
investigator). Spectrum obtained from HST web archive.}
\label{fig2}
\end{figure}
Much spectral data exists for the star, HD 44179, that powers the
fascinating Red Rectangle nebula. We here consider an apertured
line-of-sight that includes just the star, not the nebula. In such a
line-of-sight, one sees (Cohen et al. \cite{cohen}) that all Balmer lines, with the
notable exception of H($\alpha $), appear as broad absorptions against the
star's blackbody continuum. No emissions are present in the blue-green
spectral region. However, H($\alpha $) appears as a strong, narrow, emission
line centrally placed in a broad absorption. A recent STIS scan showing H($%
\alpha $) emission in HD 44179 is shown in Fig.~\ref{fig2}.

In view of the enormous variations in H($\alpha $) spectral width and
intensity observed in lines-of-sight to reddened stars of roughly the same
spectral type, a non-stellar origin for this emission is strongly suggested.
Hence - at least in the most singular cases - this emission must originate
from H atoms located outside the photosphere of the illuminating star. If
the H($\alpha $) emissions resulted from electron impact excitation
occurring in compact H II regions about the stars, one would expect also to
detect emissions at the wavelengths of H($\beta $), H($\gamma $), H($\delta $%
), etc.. At least in the case of HD 44179, these are not observed. H($\alpha $%
) emission could reasonably be produced via linear photoexcitation of H
atoms located in neutral regions bordering H II regions. However, as
discussed below, this would generally result in broadened H($\alpha $)
emission-line profiles, and cannot account for the extreme $\delta $%
-function-like profiles observed in some lines-of-sight. In essence, then,
the nonlinear photoexcitation model introduced here will attempt to explain
why some reddened early-type stars show H($\alpha $) emissions that are
comparatively broad, particularly at the base, while others (e.g. Fig.~\ref{fig1}b)
display ones that are much sharper, more intense, and more singular-looking
(i.e. unaccompanied by a significant pedestal).

  \section{Electromagnetically induced transparency (EIT)}

\begin{figure}
\resizebox{\hsize}{!}{\includegraphics{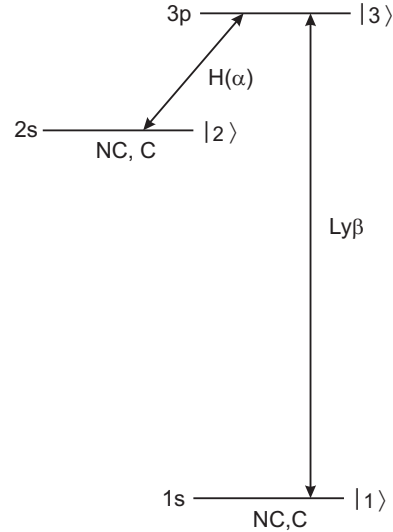}}
\caption{Energy level diagram for H-atom EIT and LWI processes discussed in
text.}
\label{fig3}
\end{figure}
  To understand the concept of LWI, one must first understand that of EIT,
since the former is built upon the latter. A good introductory review
describing some of the recent exciting developments in EIT is Harris (\cite{harris}).
The particular LWI/EIT model being considered in the present letter involves
the so-called three-level $\Lambda $ scheme shown in Fig.~\ref{fig3}. Transitions 1$%
\Longleftrightarrow $3 and 2$\Longleftrightarrow $3 are allowed; the total
radiative decay rate of quantum level $\left| 3\right\rangle $ is $\Gamma
_{3}$. Transition 1$\Longleftrightarrow $2 is assumed to be basically
forbidden. For H atoms, $\left| 2\right\rangle $ decays radiatively to $%
\left| 1\right\rangle $ via spontaneous two-photon emission at a rate $%
\approx $ 8 sec$^{-1}$. This degree of metastability qualifies the
H-atom 2s level to be level $\left| 2\right\rangle $ in a $\Lambda $-type
EIT scheme.

\begin{figure}
\resizebox{\hsize}{!}{\includegraphics{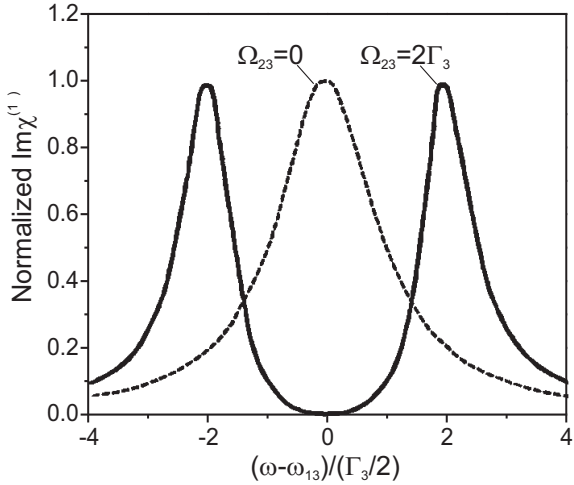}}
\caption{Absorption (Im $\chi ^{(1)}$) as a function of normalized probe
frequency offset from $\omega _{13}$, in the absence and in the presence of
a monochromatic coupling field tuned to $\omega _{23}$. From Fig. 2 of
Harris et al. (\cite{harris90}).}
\label{fig4}
\end{figure}
When a nearly monochromatic laser beam (the EIT `coupling beam'), tuned to
the center frequency $\omega _{23}$ of the 2$\Longleftrightarrow $3
transition, is applied to a gas of three-level atoms, a sharp dip in
absorption in the vicinity of the 1$\Longleftrightarrow $3 transition occurs
(Fig.~\ref{fig4}). This cancellation in the imaginary part of the linear
susceptibility (absorption is proportional to Im $\chi ^{(1)}$) can be shown
to arise from quantum interference. For the ideal case $\Gamma _{2}$=0,
there is perfect transparency at the minimum. The width of the transparency
hole varies as $\Omega _{23}$, the so-called Rabi frequency, which is itself
proportional to the square root of the laser power ($\hbar \Omega $= $\mu E$%
, where $\mu $ is the dipole matrix element and $E$ is the optical field).

If another nearly monochromatic laser beam (the EIT `probe beam'), tuned to
the center frequency  $\omega _{13}$ of the 1$\Longleftrightarrow $3
transition, is applied to the same gas of atoms, the system will quickly
adjust itself so that there is also no absorption about the line center of
the 2$\Longleftrightarrow $3 transition. This occurs independently of the
ratio of Rabi frequencies $\Omega _{13}$/$\Omega _{23}$. The two laser beams
rapidly drive all atoms of the system into a stable `coherently trapped
population state'. In this state, the wavefunction of all the atoms in the
gas becomes an antisymmetric linear combination of the unperturbed
wavefunctions of levels $\left| 1\right\rangle $ and $\left| 2\right\rangle $%
. In this state, no atoms are present in level $\left| 3\right\rangle $. One
of the earliest experiments in which coherent trapping of atomic populations
was realized is described in Gray et al. (\cite{gray}). Atoms coherently trapped by
this method are also commonly referred to as `coherently phased' atoms. The
coherently trapped state is sometimes denoted by NC, standing for
`non-coupled', because Im $\chi ^{(1)}$ $=0$ at the line centers of the two
transitions 1$\Longleftrightarrow $3 and 2$\Longleftrightarrow $3. Once all
the atoms of a gas have been driven to the NC state, light at the two
applied laser frequencies can continue to pass over the atoms without being
attenuated. There is a state orthogonal to the NC state in which the
wavefunction of the atoms is a symmetric linear combination of the
unperturbed wavefunctions of levels $\left| 1\right\rangle $ and $\left|
2\right\rangle $. Because Im  $\chi ^{(1)}\neq $ $0$ at the two allowed
transitions for atoms in this state, it is sometimes referred to as the C
(`coupled') state.

\section{Laser without inversion (LWI); pumping via hyper-Raman scattering (HRS)}

Consider a $\Lambda $-type system with all atoms maintained in the NC state
via the EIT process described above. If, through some separate means, some
of these atoms can be excited to level $\left| 3\right\rangle $, then these
atoms can fluoresce at both 1$\Longleftrightarrow $3 and 2$%
\Longleftrightarrow $3 line centers via allowed transitions to the two C
states shown in Fig.~\ref{fig3}. Since the EIT mechanism rapidly and continually
transfers all atoms from the C state to the NC state, true population
inversions can exist between level $\left| 3\right\rangle $\ and both C
states. Gain can therefore be present on these transitions, and stimulated
emission (i.e. laser action) can result. This scenario is termed a `laser
without inversion', because there are many more atoms in each of the two NC
levels than in level $\left| 3\right\rangle $. An LWI demonstrated by the
group of M.O. Scully (Padmabandu et al. \cite{padmabandu}) works by combining optical
pumping and EIT.

In considering mechanisms that could pump an H-atom LWI in space, one first
should analyze simple linear photoexcitation of NC atoms via absorption of
far-ultraviolet (FUV) continuum light from the star in the spectral vicinity
of Ly-$\beta $. Typical H-atom densities in cold, neutral clouds surrounding
bright stars are 10$^{4}$-10$^{5}$ cm$^{-3}$, implying collision rates $\sim 
$10$^{-8}$-10$^{-7}$ sec$^{-1}$. Such media should therefore be regarded as
being virtually collisionless. In a collisionless regime, FUV light in the
vicinity of Ly-$\beta $ - but offset enough from it to avoid the $\chi ^{(1)}
$ minimum (Fig.~\ref{fig4}) - will either be elastically scattered by NC atoms or
will undergo spontaneous resonant Raman scattering. In the latter process,
an atom becomes excited to the NC $\left| 2\right\rangle $ level, and a
photon slightly offset from H($\alpha $) is simultaneously emitted. However,
no excitation of level $\left| 3\right\rangle $ occurs (Loudon \cite{loudon}). Linear
photoexcitation would therefore be an ineffective pumping mechanism for LWI
in space, but it could perhaps account for the presence of spectrally broad
H($\alpha $) emissions seen from some reddened early-type stars.

\begin{figure}
\resizebox{\hsize}{!}{\includegraphics{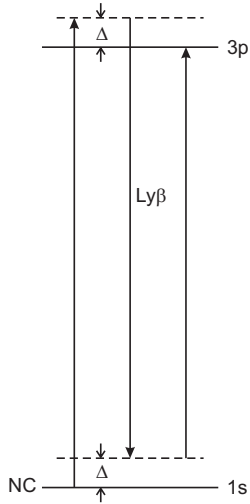}}
\caption{Resonant hyper-Raman scattering process proposed in text as the
pumping mechanism for an H-atom LWI.}
\label{fig5}
\end{figure}
Harris et al. (\cite{harris90}) were the first to show that, while EIT makes a material
system transparent to resonant laser radiation, it nonetheless allows a
large nonlinear resonant process to occur with high probability. We propose
that one such process, resonant hyper-Raman scattering (HRS), might be a
viable mechanism for pumping an H-atom LWI in space. In the unit HRS pumping
process (Fig.~\ref{fig5}), a pair of FUV photons, with frequencies lying very close
to Ly-$\beta $ but offset from it by equal amounts to high and low energies,
are absorbed from the star's blackbody continuum, a photon at Ly-$\beta $
line center is emitted, and an NC atom is excited to level $\left|
3\right\rangle $ - with all events in this energy conserving process
occurring simultaneously. This HRS mechanism would be an efficient H-atom
LWI pump, since, in the unit process, both a Ly-$\beta $ photon at line
center is generated, and an NC atom is excited to the desired 3p level. An
important feature of LWI excitation occurring via this mechanism is that the
rate of pumping is itself proportional to the intensity of the radiation
present at Ly-$\beta $ line center.

\section{H-atom LWI associated with a reddened, early-type star}

One would expect an H-atom LWI associated with a reddened early-type star to
occur in a thin shell of neutral gas located just outside an ionized
hydrogen (H II) region. For a B0 III star, the radius $R_{S}$ of such a
region would typically be $\sim $0.01 pc. At 1000 \AA , the FUV continuum
flux at $R_{S}$ propagating outward from the star would be $\sim $3 x 10$%
^{10}$ photons per cm$^{2}$ per sec per wavenumber. This flux could be
additionally enhanced several hundred times around the wavelength of Ly-$%
\beta $ via photoprocesses occurring in the H II region. It is also possible
that nonlinear elastic scattering by H atoms of the two LWI narrow-band
emissions generated would lead to additional intensity enhancement of these
emissions in the active region via the effect of decreased spatial
diffusion. This, in turn, would increase the rate of pumping.

From Fig.~\ref{fig4}, one predicts a telling signature of LWI/EIT to be extremely
narrow-band emission observed at both the H($\alpha $) and Ly-$\beta $ line
centers. The HRS pumping process should be manifest by symmetrical
absorption of continuum light around Ly-$\beta $ in a pattern that might
superficially resemble the one shown in Fig.~\ref{fig2}. (In Fig.~\ref{fig2}, however, the
apparent absorption dip has a stellar origin, as was noted above.)

An LWI/EIT mechanism can reasonably explain certain other intense, extremely
narrow emission lines recorded by astronomers, examples being the sodium
D-line doublet appearing in HD 44179, or the powerful O VI (1032 \AA , 1037 
\AA ) emission doublet that dominates the FUV spectra of symbiotic stars
such as RR Tel. In both these cases, LWI/EIT would have to occur via the
so-called `V' scheme, which incidentally is the easiest one to excite in the
presence of significant thermal Doppler broadening (Boon et al. \cite{boon}).
However, the environments existing in cold, collisionless neutral clouds
surrounding H II regions are generally favorable for all LWI/EIT schemes.

M.O. Scully has coined the word $phaseonium$ to describe an ensemble of
coherently phased atoms. An LWI based upon such a medium, he suggests, might
appropriately be termed a $phaser$. Should the occurrence of LWI/EIT in
space ever become substantiated, widespread usage of these terms in
astronomy might follow.

\end{document}